\documentclass[allclo]{FBSart}
\usepackage{amsfonts}
\usepackage{amssymb,amsmath}
\usepackage[dvips]{graphicx} 
\usepackage[section]{placeins}
\title{Two-boson Correlations in Various One-dimensional Traps%
\thanks{Article based on the presentation
by A.~Okopi\'nska at the Fifth Workshop on Critical Stability, Erice, Sicily, Received November 30, 2008; Accepted January 8, 2009}}
\author {A.~Okopi\'nska \thanks{\textit{E-mail address:}
okopin@fuw.edu.pl}, P.~ÊKo\'scik \thanks{\textit{E-mail address:}
koscik@pu.kielce.pl}}
\institute{Institute of Physics, University of Humanities and Sciences,
\'Swi\c{e}tokrzyska 15, 25-406 Kielce, Poland}
\runningauthor{A.\,Okopi\'nska and P.\,Ko\'scik}
\runningtitle{Two-boson Correlations}
\begin{document}

\maketitle
\begin{abstract}
A one-dimensional system of two trapped bosons which interact
through a contact potential is studied using the optimized
configuration interaction method. The rapid convergence of the
method is demonstrated for trapping potentials of convex and
non-convex shapes. The energy spectra, as well as natural orbitals
and their occupation numbers are determined in function of the
inter-boson interaction strength. Entanglement characteristics are
discussed in dependence on the shape of the confining potential.
\end{abstract}
\section{Introduction}
Entanglement as a measure of quantum correlations is investigated
in the hope of better understanding the structure of
strongly-coupled many-body systems. Recently there is a growing
interest in studying few-particle trapped systems, since they
became accessible in experiments with ultracold gases in optical
lattices and microtraps. The interatomic interaction can be there
considered as a contact one. By choosing the transverse
confinement much stronger than the longitudinal one, the
quasi-one-dimensional systems with an effective interaction
$g_{1D}\delta(x_{2}-x_{1})$ of an adjustable strength $g_{1D}$ may
be experimentally realized \cite{Kino}. In the Tonks--Girardeau
(TG) limit of $g_{1D}\rightarrow\infty$ the system is solvable for
arbitrary trapping potential \cite{Gir}. Theoretical consideration
of such a system evolution from weak to strong interactions is
thus of interest.

We discuss entanglement properties for a system of two bosons
interacting through a contact potential and subject to a confining
potential $V(x)$. The dimensionless Schr\"{o}dinger equation takes a
form
\begin{equation}
H\phi(x_{1},x_{2})=E \phi(x_{1},x_{2}),\label{Sch2}\end{equation}
where the Hamiltonian reads
\begin{equation}
 H= -{1\over 2}{\partial^2\over \partial x_{1}^2}-{1\over 2}{\partial^2\over \partial x_{2}^2}
 +V(x_{1})+V(x_{2})+g_{1D}\delta(x_{2}-x_{1}).
 \label{hamHOResc}
\end{equation}
%
Since the two-boson function is symmetric and may be chosen real, there exists an
orthonormal real basis $\{v_{l}\}$ such that
\begin{equation}
\phi(x_{1},x_{2})=\sum_l  k_{l}v_{l}(x_{1})
v_{l}(x_{2}),\label{funprzynat}\end{equation}
where the coefficients $k_{l}$
are real and $\sum_{l}k_{l}^2=1$. Therefore
\begin{equation}
 \int_{-\infty}^{\infty}\phi(x,x')v_{l}(x')
dx'=k_{l}v_{l}(x),\label{rrd1}
\end{equation}
which means that $v_{l}$ are eigenvectors of the two-particle function.
It may be shown that $v_{l}$ are also eigenvectors of the density
matrix, known as natural orbitals. Density matrix decomposition is
given by $ \rho(x,x^{'})=\sum\lambda_{l}
v_{l}(x)v_{l}(x^{'}),\label{densmatrixorbital}$ where the
occupancies $\lambda_{l}=k_{l}^2.$ The number of nonzero
coefficients $k_{l}$ and the distribution of their values
characterize the degree of entanglement.
\section{Optimized Configuration Interaction Method}\label{CI}
The configuration interaction method (CI) consists in choosing the
orthogonal basis set in the Rayleigh-Ritz (RR) procedure so as to
ensure proper symmetry under exchange of particles \cite{CI}. For the
two-boson system, the CI expansion reads
\begin{equation}
\phi(x_{1},x_{2})=\sum a_{ij}\psi_{ij}(x_{1},x_{2}),\label{opop}
\end{equation}
where
$\langle x_{1},x_{2}|ij\rangle=\psi_{ij}(x_{1},x_{2})\!=\!
  b_{ij}[\varphi_{i}(x_{1})\varphi_{j}(x_{2})+
  \varphi_{j}(x_{1})\varphi_{i}(x_{2})]
$ with $b_{ij}\!=\!1/2$ for $i\!=\!j$ and  $b_{ij}\!=\!1/\sqrt {2}$
for $i\!\neq \!j$. Exact diagonalization of the infinite Hamiltonian
matrix $H_{nmij}=\langle nm| H|ij\rangle$ determines the whole spectrum of
the system. Truncated matrices $[H]_{N\times N}$ allow determination of
successive approximations to the larger and larger number of states by increasing the order $N$. We use the one-particle
basis of the harmonic oscillator eigenfunctions
\begin{equation}
\varphi_{i}^{\Omega}(x)=\left(\frac{\sqrt{\Omega}}{\sqrt{\pi}2^{i}i!}
\right)^{\frac{1}{2}}H_{i}(\sqrt{\Omega} x)\exp\left[-\Omega
x^2/2\right].\label{HO}\end{equation}
Following the optimized RR
scheme~\cite{ao}, we adjust the value of the frequency $\Omega$ so
as to make stationary
the approximate sum of $N$ bound-state energies, by requiring
\begin{equation}
\frac{\delta Tr[H]_{N\times N}}{\delta \Omega}=0. \label{opt}
\end{equation}
Such a way of proceeding has been shown to improve strongly the
convergence of the RR method~\cite{ao, kos0}. The $N$th order
calculation provides approximations to many eigenstates, which
enables a direct determination of natural orbitals by representing
them in the same basis (\ref{HO}) as $v(x)=\sum p_{n}\varphi_{n}^{\Omega}(x)$.
This turns the eigenequation (\ref{rrd1}) into an algebraic problem
\begin{equation}
\begin{split}
 \sum(A_{mn}-k_{n} \delta_{mn})p_{n}&=0,\\
A_{mn}=\int
\psi_{m}^{\Omega}(x_{1})\phi(x_{1},x_{2})\psi_{n}^{\Omega}(x_{1})dx_{1}dx_{2}&=
\left\{\begin{array}{ccc}
  a_{nn}& \mbox{for} & m=n\\
  {2^{-1/2} a_{mn}}& \mbox{for} & m\neq n
\end{array}\right.
\end{split}
\end{equation}
and $a_{nm}$ are determined from diagonalization of $[H]_{N\times
N}$. By diagonalization of the matrix $[A]_{N\times N}$, the approximate
coefficients $k_{n}$
may be determined. Due to the fact that $\sum A_{nm}^2=1$, their
numerical values satisfy $\sum k_{n}^2=1$.
\section{Results}\label{results}
%

\begin{figure}[!!h]
\centerline{
\includegraphics[width=.55\textwidth]{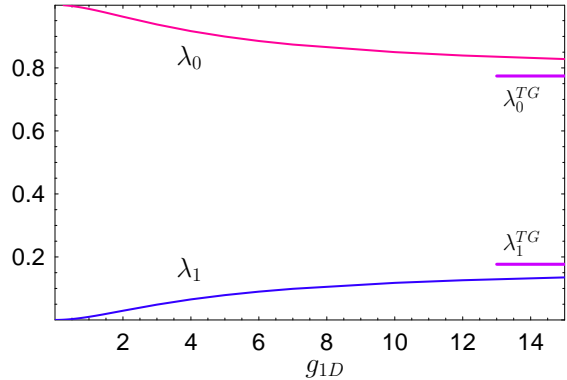}\hspace*{.8cm}}
 \caption{The occupancies $\lambda_{0}$ and $\lambda_{1}$ for a
  harmonically confined two-boson system in function of
  $g_{1D}$, their TG limits are marked by horizontal lines.}
\label{howw:beh}
\end{figure}

In the case of harmonic confinement $V(x)=m x^2/2$ and the contact
interaction, the two-particle wave function may be analytically
expressed~\cite{exact}. This allows determination of the
occupancies $\lambda_{i}=k_{i}^2$ by discretizing (\ref{rrd1}).
The two largest occupancies for the ground state are shown in Fig.
\ref{howw:beh} in function of $g_{1D}$. The state is non-entangled
($\lambda_{0}=1$) only if the bosons do not interact. The weakly
entangled "condensed" state with only one orbital significantly
occupied is realized at very weak interactions, $g_{1D}\lesssim
0.1$. With increasing $g_{1D}$, the entanglement grows, which
shows up in the increase of $\lambda_{1}$ at the cost of
$\lambda_{0}$. The occupancies monotonically approach their TG
limits $\lambda_{0}^{TG} \approx 0.7745$ and $\lambda_{1}^{TG}
\approx 0.1765$.

\begin{figure}[!t]
\begin{center}
\includegraphics[width=1.00\textwidth]{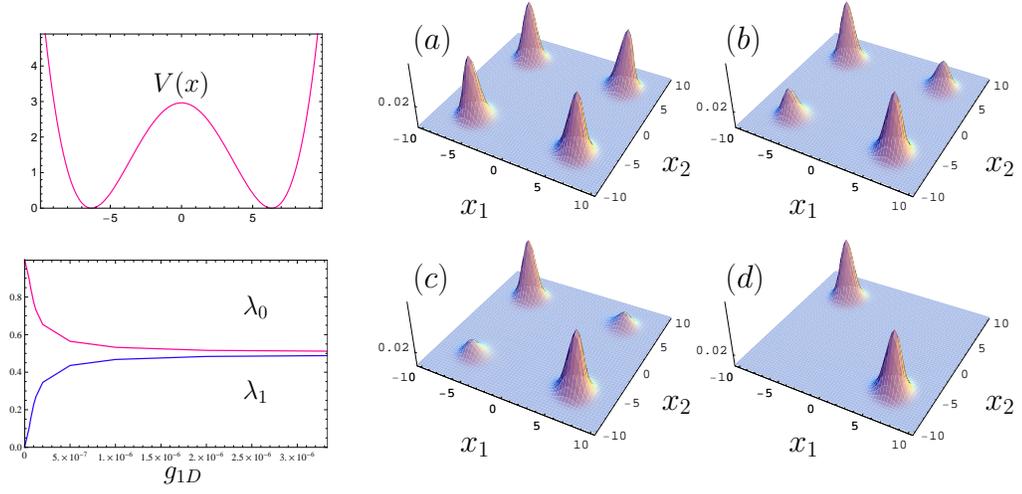}
\end{center}
 \caption{\label{1wpor322:beh}
 Double-well potential (upper left), the occupancies $\lambda_{0}$ and $\lambda_{1}$ in function of
  $g_{1D}$ (lower left) and two-boson densities (right) for (a) $g_{1D}=0 $,
  (b) $g_{1D}=2.5\cdot 10^{-8}$, (c) $g_{1D}=5\cdot 10^{-8}$, (d) $g_{1D}=10^{-6}$  }
\end{figure}

\begin{figure}[!!h]
\begin{center}
\includegraphics[width=1.00\textwidth]{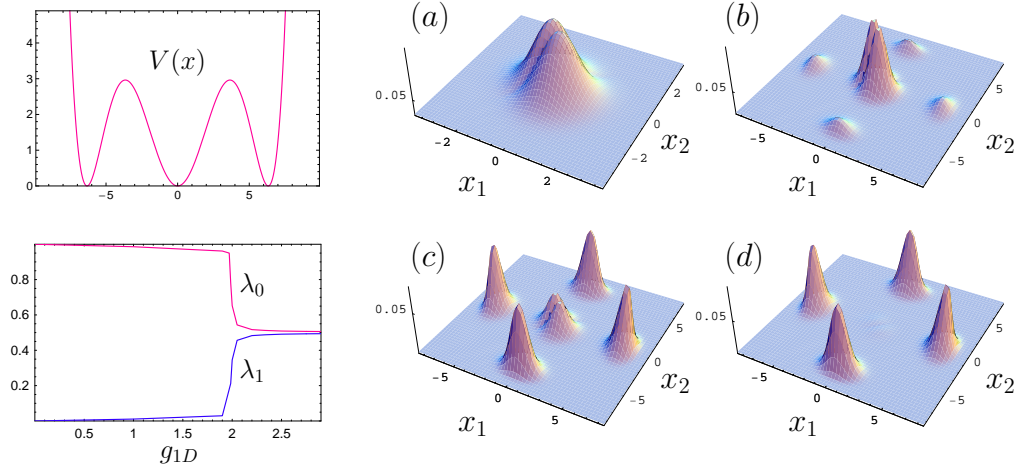}
\end{center}
 \caption{\label{1wpor322e:beh}Same as Fig.\ref{1wpor322:beh} but
 for the triple-well potential. The two-boson densities (right) for (a) $g_{1D}=1$,
  (b) $g_{1D}=1.97$, (c) $g_{1D}=1.985$, (d) $g_{1D}=2.05$ }
\end{figure}

Entanglement properties in the case of multi-well potentials are
markedly different. Using the
optimized RR method, we calculated the natural orbital occupancies
of ground states in double-well potential $V_{2{\rm
well}}({x})={2\over 27 a}(1- a x^2)^2$ and triple-well potential
$V_{3{\rm well}}(x)={1\over 2}x^2 -a x^4 + {a^2\over 2} x^6$. The potentials have minima of the same depth
and the maxima of the same height, controlled by the parameter $a$.
The results for $a=0.025$ are plotted in Figs.~\ref{1wpor322:beh} and
\ref{1wpor322e:beh}, where the upper left presents the shapes of the
potentials, and the lower left, the two largest occupancies
$\lambda_{0}$ and $\lambda_{1}$ in function of $g_{1D}$. For
$g_{1D}=0$, the ground state is non-entangled, as $\lambda_{0}=1$.
With increasing interactions, $\lambda_{0}$ decreases and
$\lambda_{1}$ grows, monotonically approaching the TG limit of
non-entangled "fragmented" state,
$\lambda_{0}^{TG}=\lambda_{1}^{TG}= 0.5$. The critical value
$g_{1D}^{cr}$, above which $\lambda_{0}\approx\lambda_{1}$, is much
larger for the triple-well potential than for the double-well one.
The dependence of the two-boson density on $g_{1D}$ for the
double-well potential is shown on the right of
Fig.~\ref{1wpor322:beh}.
For noninteracting bosons, the probability
of both being in different wells is the same as being in the same
well. With increasing $g_{1D}$, the probability of finding the bosons in
the same well quickly decreases and above $g_{1D}^{cr}$ the state is
almost fragmented. In the triple-well case (right of
Fig.~\ref{1wpor322e:beh}) the particles live in the middle well, only
above $g_{1D}^{cr}$ the probability of finding a particle in an
external well becomes considerable. In the TG limit of non-entangled
``fragmented" state, one particle is localized in the middle and the
other in one of external wells.

\section{Conclusion}
The optimized CI method proves very effective in determining the
spectrum and the natural orbitals of the two-particle confined
systems.

\begin{small}

\end{small}


\begin{thebibliography}{9}
\bibitem{Kino} Kinoshita, T., Wenger, T., Weiss, D. S.: Science {\bf 305}, 1125
(2004).
\bibitem{Gir} Girardeau, M.: J.Math.Phys. {\bf 1}, 516 (1960)
\bibitem{CI} Helgaker, T., J\o rgensen P., Olsen J.: Molecular Electronic-
Structure Theory (Wiley, Chichester, 2000).
\bibitem{ao} Okopi\'nska, A.: Phys.Rev. {\bf D36}, 1273 (1987)
\bibitem{kos0} Ko\'scik, P., Okopi\'nska, A.: J. Phys. A: Math. Theor. {\bf 40}, 10851 (2007)
\bibitem{exact} Busch, T., et al.:
Found. Phys. {\bf 28}, 549 (1998)
\end{thebibliography}
\end{document}